# Encircling exceptional points in a Riemann sphere for efficient asymmetric polarization-locked devices

Aodong Li[1], and Lin Chen[1,*]

[1] Wuhan National Laboratory for Optoelectronics, Huazhong University of Science and Technology, Wuhan 430074, China

* Correspondence authors: chen.lin@hust.edu.cn

**Abstract**

Dynamically encircling exceptional points (EPs) in two-dimensional Hamiltonian parameter space has enabled intriguing chiral dynamics in which the final state of the system depends on the encircling direction. Here, we show that full Hamiltonian parameter space can be described in a Riemann sphere, and those points on the parameter space boundary with the eigenstates of the system Hamiltonian being convergent, converge to the north vertex. We present that encircling one EP on the Riemann sphere leads to chiral response, and a continuous encircling trajectory passing through the north vertex can realize near-unity asymmetrical transmission. An asymmetric polarization-locked devices are designed by mapping the encircling path onto the L-shaped silicon waveguides. We experimentally demonstrate near 100% asymmetrical polarization conversion efficiency between TE and TM modes with the mode crosstalk below -20 dB at 1550 nm. Our results bring the study of EP-associated dynamics into the realm of highly-efficient asymmetrical polarization conversion and power up new application opportunities for EP physics.

In a non-Hermitian system with parity-time (PT) symmetry, two or more eigenvalues and eigenstates are degenerate simultaneously at exceptional points (EPs). A variety of intriguing phenomena, such as sensing enhancement[1,2], unidirectional invisibility[3,4], topological light control[5,6], and asymmetric mode conversion[7-15], arising from the unique topological features of EPs, have been theoretically predicted and experimentally demonstrated, and have attracted ever growing interest in the fields of optics[1,6,7,16-24], acoustics[25,26], thermodynamics[27], electronics[28,29] and quantum mechanics[30]. The EP in an optical system can be well observed in coupled waveguides with tailored gain and loss, which has led to loss induced optical transparency[16]. The Berry phase is $\pi$ when the system Hamiltonian encircles an EP twice, indicating a self-intersecting Riemann surface around EP[31].

Recent studies have found that dynamically encircling an EP can enable asymmetric response, that is, encircling an EP in a clockwise (CW) or anticlockwise (ACW) loop leads to different final states, regardless of the starting states. Such chiral behavior has been theoretically and experimentally demonstrated by mapping the required Hamiltonian parameters onto PT symmetric arrangements of waveguides[7,19]. Most of the previous studies have been conducted to realize asymmetric conversion of waveguide modes in optical waveguides, but failed to achieve high transmission efficiencies arising from path-dependent loss[7,11,12,19,32,33]. To solve the ultra-low loss issue, the authors have recently proposed a discontinuous encircling of an EP via a path connecting the convergent eigenstates of the system as its Hamiltonian parameters approach infinity[15]. Two attempts have been made to utilize EP encircling strategy to produce a single desired output light polarization, regardless of the input light polarization state[9,34]. Such a polarization-locked device is particularly important in an optical communication circuit, such as polarization-dependent detection. The first work was numerically proposed by use of a single GaAs-AlGaAs waveguide, where an optical gain was assumed to be provided in a specific waveguide region. Such a scheme faces significant challenge in practical experimental implementation, and has not been experimentally demonstrated yet[9]. Quite recently, encircling an EP in fiber-based systems has enabled the realization of polarization-dependent output. However, it merely works under the condition that an optical pulse is input, and real-time asymmetrical polarization conversion is unrealizable in principle[34].

In this article, we show that full Hamiltonian parameter space can be described in a Riemann sphere, where the infinite points on the boundary space, associated with the same eigenstates, converge to the north vertex. Encircling an EP splits the Riemann sphere into two surfaces, with a Berry phase being of $\pi$. A continuous trajectory passing through the north vertex promises highly-efficient asymmetrical transmission. Experimental results by mapping this trajectory onto a suitably designed waveguide of L-shaped cross-sections suggest near-unity asymmetrical conversion efficiency between TE and TM modes with the mode crosstalk below -20 dB at 1550 nm. It should be mentioned that, the device is referred to asymmetrical polarization-locked devices (APLDs) since the output polarization state is locked and irrespective of the input polarization states.

A system state, $|\psi\rangle$, that evolves along $z$ direction abides by the Schrödinger-type equation $i\partial/\partial z|\psi\rangle = H|\psi\rangle$, with the second-order Hamiltonian $H$ being written as

$$H = \begin{bmatrix} \beta & \kappa \\ \kappa & -\beta + i\gamma \end{bmatrix} \tag{1}$$

where $\beta$, $\gamma$ and $\kappa$ represent the degree of detuning, relative gain/loss rate, and coupling strength of the system, respectively. The two eigenvalues are $E_{1,2} = i\gamma/2 \pm \sqrt{\kappa^2 + (\beta - i\gamma/2)^2}$, and the associated eigenstates are $|\psi_{1,2}\rangle = \left[\sqrt{1 \pm K}, \pm \sqrt{1 \mp K}\right]^T / \sqrt{2}$, with $K = (\beta - i\gamma/2)/\sqrt{\kappa^2 + (\beta - i\gamma/2)^2}$. In the two-dimensional (2-D) parameter space $\vec{R} = (\beta/\kappa, \gamma/\kappa)$, the EPs are located at $\vec{R} = (0, 2)$ and $\vec{R} = (0, -2)$. When $H$ approaches the infinite points on the parameter space boundary ($\beta/\kappa \to \pm\infty$ and/or $\gamma/\kappa \to \pm\infty$), the two eigenstates converge to $|\psi_1\rangle = [1, 0]^T$ and $|\psi_2\rangle = [0, 1]^T$, corresponding to the eigenvalues $E_1$ and $E_2$, respectively. The energy spectra form a self-intersecting Riemann surface in the parameter space $\vec{R}$, and the topological features can be expressed as a Berry phase of $\Phi = \pi$, when an EP is encircled twice in a quasi-dynamic approach[35]. The Berry phase can be retrieved with $\Phi = i \int_{\vec{R}_1}^{\vec{R}_2} \langle \psi(\vec{R}) | \nabla_{\vec{R}} \psi(\vec{R}) \rangle \cdot d\vec{R}$, where the Hamiltonian adiabatically evolves from $\vec{R}_1$ to $\vec{R}_2$ ($\vec{R}_1$ and $\vec{R}_2$ are arbitrarily selected two points on the parameter space)[36]. It is the reason that $\psi(\vec{R})$ converges to a constant value, associated with $\left|\nabla_{\vec{R}}\psi(\vec{R})\right\rangle = 0$, that renders $\Phi$ equal to zero as the Hamiltonian evolves along the parameter space boundary [Fig. 1(a)].

Here, we propose to use a Riemann sphere to represent the full parameter space, where the infinite points on the parameter space converge to the north vertex, as schematically shown in Fig. 1(b). Consequently, the infinite points that can merely be schematically described in the 2-D parameter space can be explicitly defined in a Riemann sphere. The north vertex, associated with those infinite points in the 2-D parameter space, does not contribute to the Berry phase. Here, the radius of the Riemann sphere is assumed to be 2 and the mapping relation from the 2-D parameter space to the Riemann sphere is $\beta/\kappa = 2\cot(\varphi_1/2)\cos(\varphi_2)$ and $\gamma/\kappa = 2\cot(\varphi_1/2)\sin(\varphi_2)$, where $\varphi_1$ is the zenith and $\varphi_2$ is the azimuth as denoted in Fig. 2(b). Encircling two EPs and no EP both result in a Berry phase of $\Phi = 0$. In both cases, the Riemann sphere is split into two surfaces, in which one contains two EPs and another contains no EP [Figs. 1(c) and 1(d)]. In contrast, the encircling trajectory including one EP splits the Riemann sphere into two surfaces that each contains one EP [Fig. 1(e)], associated with a Berry phase of $\Phi = \pi$.

A general encircling trajectory around an EP for asymmetrical transmission is depicted with yellow curves as shown in Fig. 2(a). Since it does not pass through the north vertex of the Riemann sphere, associated with path-dependent loss along the whole path ($\gamma/\kappa > 0$), renders the system state suffers from low transmittance [Fig. 2(a)]. The loss difference of eigenstates causes the occurrence of nonadiabatic transitions (NATs), which leads to asymmetrical response[10,11,37]. A specific trajectory passing through the north vertex, described by the black curves in Fig. 2(a), does not make the system undergo any loss except at the north vertex, and hence enables high-efficiency asymmetrical transmission. This specific trajectory can be mapped onto a suitably designed silicon waveguide for asymmetrical polarization conversion, as schematically depicted in Fig. 2(b). The right waveguide supports two orthogonal eigenmodes with different polarization directions, that is R1 and R2 modes with the effective refractive indices being of $n_{R1}$ and $n_{R2}$ ($n_{R1} < n_{R2}$), respectively. $\theta$ stands for the angle between the polarization direction of R1 mode and $+45°$ axis [Fig. 2(c)]. Here, the eigenmode in the right waveguide can be expressed as $|\psi\rangle = [A_1, A_2]^T$, in which $A_1$ and $A_2$ indicate the projection components of normalized electric field amplitude along $+45°$ and $-45°$ axis, respectively. Consequently, the Hamiltonian parameters in Eq.1 can be expressed as $\beta = k_0(n_{R1} - n_{R2})(\sin^2\theta - \cos^2\theta)/2$, $\kappa = k_0(n_{R1} - n_{R2})\sin\theta\cos\theta$. $\gamma$ is the loss rate exerted on the projection components of electric field along $-45°$ direction, and $k_0 = 2\pi/\lambda$ is the wavenumber with $\lambda$ being of light wavelength (see the Supplemental Material, Note 1).

At the starting point A, R1 mode is transverse-magnetic-like (TM) polarized with $\theta = 45°$, corresponding to $\beta/\kappa = 0$ [Fig. 2(d)]. Between points A and B, $\theta$ changes monotonously from $45°$ to $0°$, making $\beta/\kappa$ reduce from 0 to $-\infty$, which is associated with the evolution path from A to B along the longitude on the Riemann sphere [Fig. 2(b)]. Between points B and C, the right waveguide is designed to be axisymmetric along $+45°$ axis, so that R1 mode polarizes along $+45°$ direction with $\theta = 0°$, resulting in $\kappa = 0$ and $\beta/\kappa = \infty$ [Fig. 2(d)]. The left waveguide between B and C is designed to absorb the eigenmode polarized along $-45°$ direction in the right waveguide via adiabatic coupling, associated with $\gamma/\kappa \to +\infty$ for $\gamma > 0$. The whole evolution path between B and C converges to the north vertex, M, on the Riemann sphere. Between points C and D, $\beta/\kappa$ is varied monotonously from $+\infty$ to 0 as $\theta$ reduces from $0°$ to $-45°$, and the Hamiltonian evolves from C to D (the starting point A) along the longitude on the Riemann sphere.

Figures 3(a-d) show the evolution trajectory of the system state on the Riemann surfaces for the CW and ACW loops around the EP, where the CW and ACW loops

indicates light propagates along $+z$ and $-z$ directions in the right waveguide, respectively. The color of the Riemann surfaces represents the imaginary part of Im(E) of the energy spectra. For the CW loop with $[1,1]^T$ input, associated with TE mode, $[1,1]^T$ evolves to $[0,1]^T$ when the Hamiltonian varies slowly from A to B [Fig. 3(a)]. Meanwhile, $[1,0]^T$ is triggered at B, owing to the fact that rigorous adiabaticity cannot be fully guaranteed. During the NAT process from B to C, $[0,1]^T$ is dissipated and $[1,0]^T$ is retained, rendering the dominant eigenstate of the system switches from $[0,1]^T$ to $[1,0]^T$. Finally, the Hamiltonian returns to the destination D with $[1,1]^T$, and the corresponding output mode is still TE mode. For the CW loop with $[1,-1]^T$ input [Fig. 3(b)], associated with TM mode, the dominate eigenstate during the evolution is always lossless because the NAT process does not occur, and the final output state is eventually $[1,1]^T$, associated with TE mode. This indicates that the output mode for the CW direction is locked to TE mode, irrespective of the input mode. For the ACW loop [Figs. 3(c, d)], the output state is locked to $[1,-1]^T$, associated with TM mode. It should be noted that, the two sheets of the Riemann surfaces are self-intersecting, and the evolution path between B and C is vertical as the slope at B and C is infinite on the Riemann surface. On the Riemann surface, the NAT occurs between B to C, during which the evolution path switches from the blue branch to the red branch [Figs. 3(a, d)].

    The waveguide structural parameters are carefully designed to fulfil the required evolution path as described in Fig. 3. Figure 4(a) shows the dependence of $\theta$ on $w_{R1}$ and $w_{R2}$, along the waveguide propagation direction from A to D. From points A (C) to B (D), $\theta$ changes from $+45°$ (0) to 0 ($-45°$). Figures 4(b) and 4(c) show the effective refractive index of the R1 and R2 modes along the propagation direction. The length between A and B (C and D) is optimized to balance the portion of the triggered eigenmode and output mode purity. If the length is too large, the triggered eigenmode will be completely suppressed, and hence can not be the dominant eigenmode after the

NAT process. As a result, no energy outputs for the CW (ACW) loop with the TE (TM) mode. If the length is too short, the purity of the expected output mode is low. Finally, the length between A and B (C and D) is optimized to be 7 μm (6 μm).

The polarization-dependent loss is introduced by carefully designing a L-shaped waveguide on the left side that completely takes the optical energy of the R2 mode out of the right waveguide via coupling without affecting the R1 mode. The left waveguide supports L1 and L2 modes with a typical mode field distributions being shown in Fig. 4(d). The structural parameters ($w_{L1}$, $w_{L2}$) are optimized to ensure that the coupling merely occurs between the L1 and R2 modes as $n_{L1}$ merely intersects $n_{R2}$ in the interval between B and C, without affecting the R1 mode [Fig. 4(d)]. Here we have used the adiabatic width parameters ($w_{L1}$, $w_{L2}$) to achieve adiabatic coupling so that R2 mode is completely transferred to L1 mode. The left waveguide at B and C is placed as far as 600 nm from the right waveguide to ensure that the input R2 mode at B is completely converted to L1 mode at C. The gap separation between the two waveguides, $d$, should be quickly reduced to shorten the device length between B and C. A minimum $d$ =100 nm is chosen by considering the fact that an even smaller gap separation causes fabrication error. The practical structural parameters ($w_{L1}$, $w_{L2}$, $d$) used for the device are denoted by the red lines shown in Figs. 4(e) and 4(f). In addition, the device length between B and C is set as long as 120 μm to ensure adiabatic evolution. The detailed structural parameters for the whole device can be found in Note 2 of the Supplemental Material.

For the CW (ACW) loop with TE (TM) mode input to the right waveguide at A (D), TE (TM) mode evolves to R2 mode that polarizes along −45° direction at B (C) [Fig. 5(a)]. Meanwhile, R1 mode that polarizes along +45° direction is triggered because the length between A and B (D and C) is limited and adiabatic evolution is not strictly fulfilled. The device length between B and C is long enough so that R2 mode is fully absorbed by the left waveguide and the triggered R1 mode is retained, i.e., NAT occurs. Finally, R1 mode evolves to TE (TM) mode from C to D (B to A). For the CW (ACW) loop with TM (TE) mode input to the right waveguide at A (D), it evolves to R1 mode at B (C). The triggered R2 mode is absorbed between B and C, and R1 mode eventually evolves to TE (TM) mode from C to D (B to A).

The double silicon waveguides in Fig. 2(b) were fabricated by a combination of three-step electron-beam lithography (EBL), inductively coupled plasma (ICP) etching, electron-beam evaporation (EBE), and plasma-enhanced chemical vapor deposition (PECVD). The first-step EBL and EBE were used to form the Au marks on 340 nm silicon-on-insulator (SOI) platform for alignment. The second-step EBL and ICP were employed to define the partially-etched waveguide pattern with an etching depth of 120 μm. The third-step EBL and ICP were employed to define the fully-etched waveguide pattern with an etching depth of 340 μm. Finally, a 1-μm thick $SiO_2$ cladding layer is deposited to cover the entire device by PECVD. More fabrication details are provided in the Supplementary Note 3.

Figures 5(b) show the optical image of the fabricated devices with grating couplers (GCs), polarization beam splitter and rotators (PBRs) for measurement, in which the Scanning Electron Microscope imaging for a partial coupling region is shown in the right panel. The measured transmission efficiencies for TE and TM mode input are shown in Fig. 5(c) and 5(d), respectively (see Note 4 of the Supplemental Material for detailed measurement scheme). $T_{P \to Q}$ ($T_{P \leftarrow Q}$) represents the transmission efficiency of the mode Q from the output port as P mode inputs from the left (right) port (P or Q mode refer to TE or TM mode). For the left-side injection, we have $T_{TE \to TE} \gg T_{TE \to TM}$ and $T_{TM \to TE} \gg T_{TM \to TM}$, indicating that the output mode is locked to the TE mode. As for the right-side injection, we have $T_{TM \leftarrow TE} \gg T_{TE \leftarrow TE}$ and $T_{TM \leftarrow TM} \gg T_{TE \leftarrow TM}$, indicating that the output mode is locked to the TM mode. It can be observed that, $T_{TM \to TE}$ and $T_{TM \leftarrow TE}$ are close to 100% around 1550 nm. The mode crosstalk is defined as the energy ratio of the undesired mode to total output[38]. The mode crosstalk of $TM \to TE$, $TM \leftarrow TE$, and $TE \to TE$ are all below -20 dB, while that for $TM \leftarrow TM$ is below -10 dB. The experimental results well validate the asymmetrical polarization-locked performance.

It has been recently reported that asymmetric transmission for spatial light can be realized by use of a number of optical components without use of the EP encircling strategy presented here[39]. The drawback of such scheme may be the limited working bandwidth that is restricted by one of the optical components that has the smallest bandwidth. The resultant asymmetrical transmission devices enabled by EP encircling own the advantage of large working bandwidth due to adiabatic parameter evolution and robust Riemann surface of energy spectra[7,15,19]. Although the measured bandwidth for the APLD is only several dozens of nanometers, arising from the bandwidth of the laser source, the simulated results demonstrate the practical bandwidth reaches as large as 1450-1700 nm (see the simulation results in Note 5 of the Supplemental Material). Compared with the previous on-chip scheme by use of optical waveguides with the assumption of gain provided on a specific device region, our proposal can facilitate the realization of asymmetrical polarization conversion practically on a chip[9]. The asymmetrical polarization-locked output is highly promising for polarization-locked laser[40]. Future steps might be taken to enable optical isolation by linking the presented EP encircling strategy with nonlinear materials[32].

In conclusion, we have shown that dynamically encircling an EP on the Riemann sphere can enable asymmetrical transmission, and a continuous encircling trajectory passing through the north vertex can realize near-unity asymmetrical transmission. With well design of double-coupled silicon waveguides to follow the encircling path, near 100% asymmetrical polarization conversion between TE and TM modes with the mode crosstalk below -20 dB at 1550 nm is achievable. These results facilitate the practical realization of highly efficient asymmetrical polarization conversion on a chip without gain, and promise new opportunities for optical devices and applications that requires asymmetrical polarization-locked output.

**Acknowledgement**

This work is supported by National Natural Science Foundation of China (Grant Nos. 11674118, 12074137), State Key Laboratory of Artificial Microstructure & Mesoscopic Physics (Peking University), and State Key Laboratory of Advanced Technology for Materials Synthesis and Processing (Wuhan University of Technology). We thank Pan Li in the Center of Micro-Fabrication and Characterization (CMFC) of WNLO for the support in plasma enhanced chemical vapor deposition, and the Center for Nanoscale Characterization & Devices (CNCD), WNLO, HUST for the support in SEM measurement.

**Author Contributions**

A.L., and L.C. conceived the idea and initiated the work. L.C. guided the project. A.L. developed the theoretical framework, performed the numerical simulations, and performed the measurements. A.L., and L.C. wrote the manuscript and all authors reviewed the manuscript.

**Data availability**

The authors declare that all data supporting the findings of this study are available from the corresponding authors upon reasonable request.

**Conflict of interest**

The authors declare that they have no conflict of interest.

# Figures

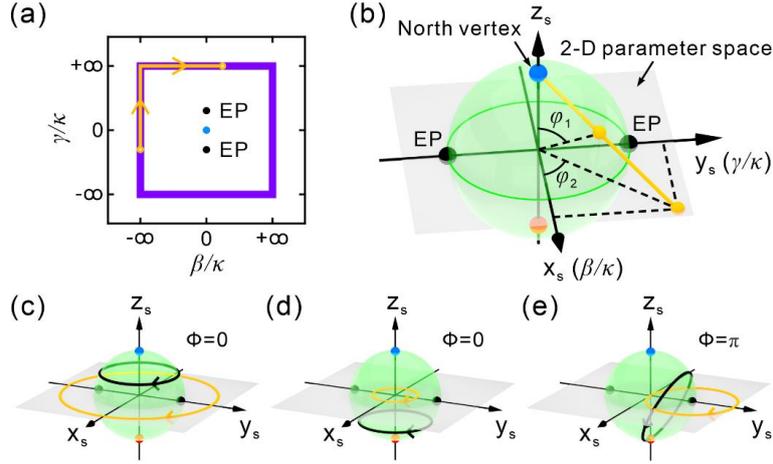

FIG. 1. (a) The Hamiltonian parameter space described by $\vec{R}=(\beta/\kappa,\gamma/\kappa)$. The yellow line shows an evolution path along the parameter space boundary that is represented by the purple line. (b) The Riemann sphere used for describing the Hamiltonian parameter space. The center of the sphere coincides with the origin of the 2-D parameter space as denoted in (a). The two EPs are included in the equator that is located in the 2-D parameter space. The yellow line is the straight line that connects the north vertex and the point on the 2-Dparameter space. Every mapping point on the Riemann sphere is the intersection point of the yellow line and the Riemann sphere. On the Riemann sphere, $x_s = 2\sin(\varphi_1)\cos(\varphi_2)$ and $z_s = 2\cos(\varphi_1)$ under the assumption that the radius of the Riemann sphere is 2. (c-e) The evolution loops on the Riemann sphere (black curves) and in 2-D parameter space (yellow curves) when encircling (c) two EPs, (d) no EP, (e) one EP.

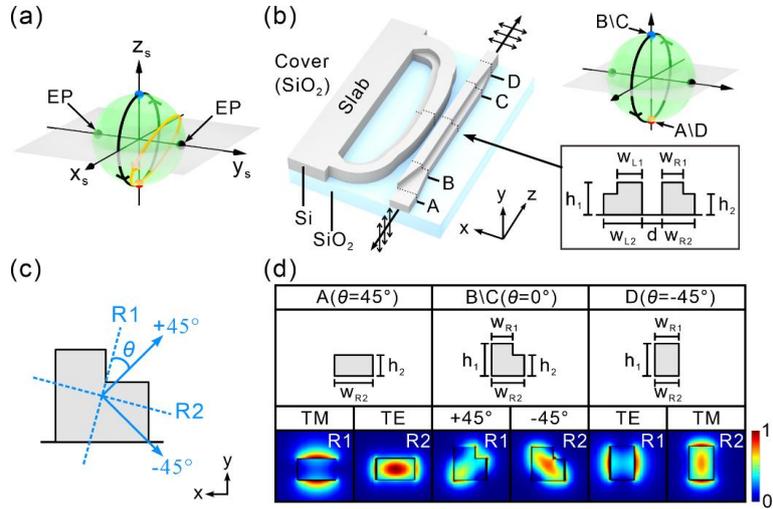

FIG. 2. (a) Two encircling loops around the EP on the Riemann sphere, in which one passes through the north vertex (black curves), and the other one does not. (b) The APLD made of the silicon waveguides in which the associated EP encircling path contains north vertex. The full height of the L-shaped waveguide is $h_1$=340 nm, and the height of the lower waveguide is $h_2$=220 nm. The width of the upper waveguide and the full waveguide is $w_{L1}$ ($w_{R1}$) and $w_{L2}$ ($w_{R2}$) for the left (right) waveguide, respectively. The gap separation between the two waveguides is $d$. (c) The polarization directions for R1 and R2 modes. (d) The cross sections of the right waveguide at A, B\C, and D, and their associated electric field intensity distributions for the R1 and R2 eigenmodes.

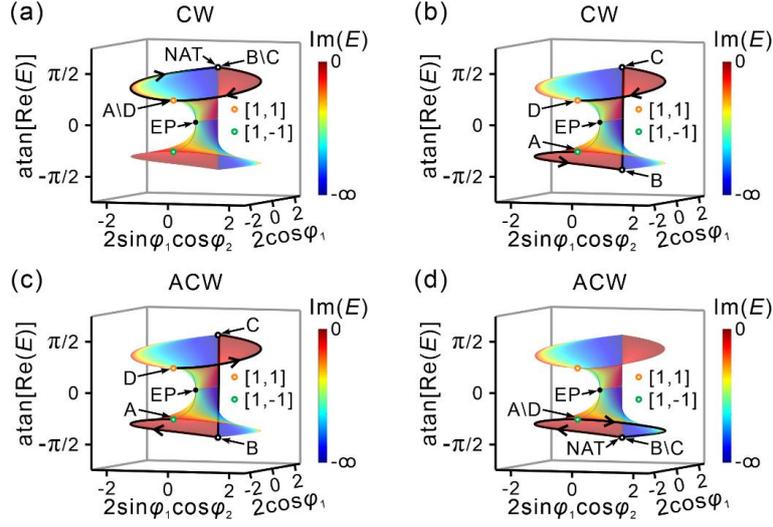

FIG. 3. (a),(b) The CW loops on the Riemann surfaces formed by the real part of the eigenvalues of $H$ when the input state is (a) $[1,1]^T$ and (b) $[1,-1]^T$. The color of the surface indicates the imaginary part of the eigenvalues of $H$, in which a smaller imaginary part indicates a larger loss. (c),(d) The ACW loops when the input state is (c) $[1,1]^T$ and (d) $[1,-1]^T$.

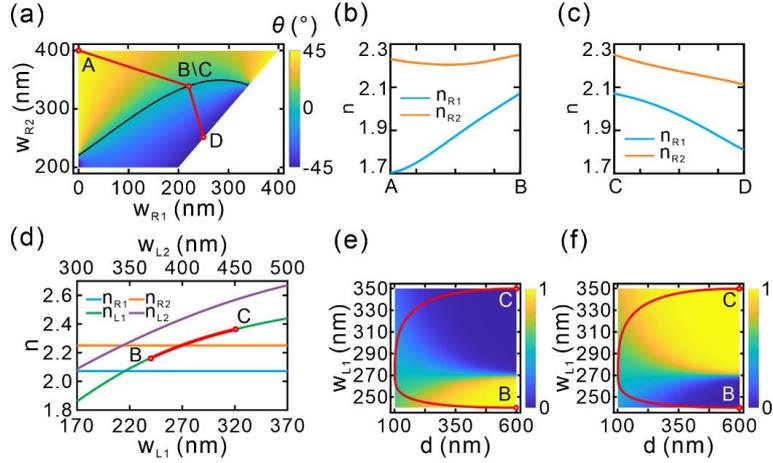

FIG. 4. (a) $\theta$ versus $w_{R1}$ and $w_{R2}$ along the waveguide propagation direction from A to D. (b),(c) The effective refractive index of the R1 and R2 modes from (b) A to B and (c) C to D. (d) The effective refractive index of the R1, R2, L1 and L2 modes between B and C as a function of $w_{L1}$ and $w_{L2}$. Here, $w_{L2} - w_{L1}$ is fixed at $130\ nm$. (e),(f) Normalized optical power of the eigenmode in the right (e) and left (f) waveguides versus $w_{L1}$. The red line depicts the geometrical parameters used between B and C.

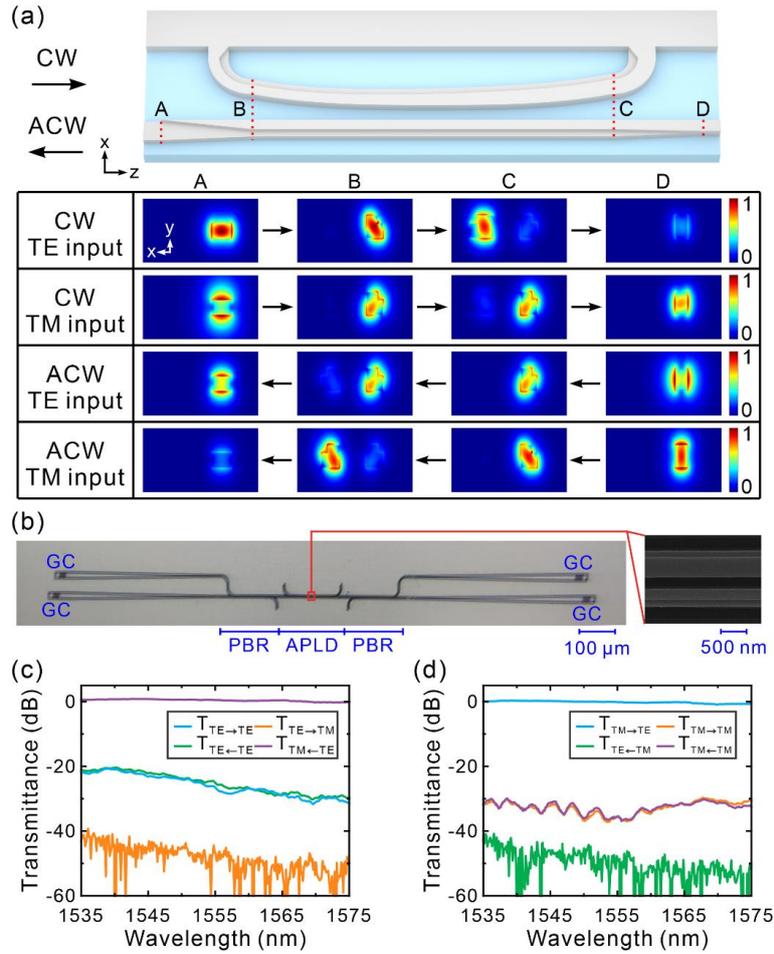

FIG. 5. (a) The cross-sectional electric field intensity distributions at A, B, C, and D. (b) The optical microscope images of the fabricated APLD, where the L-shaped waveguides are partially depicted by the SEM image on the right panel. The grating coupler (GC) is used to couple the TE mode from the waveguide into the fiber or couple the light waves from the fiber into the TE mode in the waveguide. The polarization beam splitter and rotator (PBR) is used to input TE or TM mode, or split the TE and TM modes for independent detection. (c),(d) Measured transmittance spectra over 1535−1575 nm wavelength range as (c) TE and (d) TM modes are injected into the APLD, respectively.

**References**

1. Hodaei, H. *et al.* Enhanced sensitivity at higher-order exceptional points. *Nature* **548**, 187-191, doi:10.1038/nature23280 (2017).
2. Chen, W., Kaya Ozdemir, S., Zhao, G., Wiersig, J. & Yang, L. Exceptional points enhance sensing in an optical microcavity. *Nature* **548**, 192-196, doi:10.1038/nature23281 (2017).
3. Lin, Z. *et al.* Unidirectional Invisibility Induced by PT-Symmetric Periodic Structures. *Physical Review Letters* **106**, 213901, doi:10.1103/PhysRevLett.106.213901 (2011).
4. Feng, L. *et al.* Experimental demonstration of a unidirectional reflectionless parity-time metamaterial at optical frequencies. *Nat Mater* **12**, 108-113, doi:10.1038/nmat3495 (2013).
5. Zhao, H. *et al.* Non-Hermitian topological light steering. *Science* **365**, 1163-1166, doi:10.1126/science.aay1064 (2019).
6. Song, W. *et al.* Breakup and Recovery of Topological Zero Modes in Finite Non-Hermitian Optical Lattices. *Physical Review Letters* **123**, doi:10.1103/PhysRevLett.123.165701 (2019).
7. Doppler, J. *et al.* Dynamically encircling an exceptional point for asymmetric mode switching. *Nature* **537**, 76-79, doi:10.1038/nature18605 (2016).
8. Ghosh, S. N. & Chong, Y. D. Exceptional points and asymmetric mode conversion in quasi-guided dual-mode optical waveguides. *Sci Rep* **6**, 19837, doi:10.1038/srep19837 (2016).
9. Hassan, A. U., Zhen, B., Soljacic, M., Khajavikhan, M. & Christodoulides, D. N. Dynamically Encircling Exceptional Points: Exact Evolution and Polarization State Conversion. *Physical Review Letters* **118**, 093002, doi:10.1103/PhysRevLett.118.093002 (2017).
10. XL Zhang, T. J., HB Sun, CT Chan. Dynamically encircling an exceptional point in anti-PT-symmetric systems asymmetric mode switching for symmetry-broken states. *arXiv preprint arXiv:1806.07649* (2018).
11. Zhang, X.-L., Wang, S., Hou, B. & Chan, C. T. Dynamically Encircling Exceptional Points: In situ Control of Encircling Loops and the Role of the Starting Point. *Physical Review X* **8**, 021066, doi:10.1103/PhysRevX.8.021066 (2018).
12. Zhang, X.-L. & Chan, C. T. Dynamically encircling exceptional points in a three-mode waveguide system. *Communications Physics* **2**, 63, doi:10.1038/s42005-019-0171-3 (2019).
13. Zhang, X.-L., Jiang, T. & Chan, C. T. Dynamically encircling an exceptional point in anti-parity-time symmetric systems: asymmetric mode switching for symmetry-broken modes. *Light: Science & Applications* **8**, 88, doi:10.1038/s41377-019-0200-8 (2019).
14. Liu, Q. *et al.* Efficient Mode Transfer on a Compact Silicon Chip by Encircling Moving Exceptional Points. *Physical Review Letters* **124**, 153903, doi:10.1103/PhysRevLett.124.153903 (2020).
15. Li, A. *et al.* Hamiltonian Hopping for Efficient Chiral Mode Switching in

Encircling Exceptional Points. *Physical Review Letters* **125**, 187403, doi:10.1103/PhysRevLett.125.187403 (2020).

16. Guo, A. *et al.* Observation of PT-symmetry breaking in complex optical potentials. *Physical Review Letters* **103**, 093902, doi:10.1103/PhysRevLett.103.093902 (2009).

17. Lawrence, M. *et al.* Manifestation of PT symmetry breaking in polarization space with terahertz metasurfaces. *Physical Review Letters* **113**, 093901, doi:10.1103/PhysRevLett.113.093901 (2014).

18. Peng, B. *et al.* Chiral modes and directional lasing at exceptional points. *Proceedings of the National Academy of Sciences* **113**, 6845-6850, doi:10.1073/pnas.1603318113 (2016).

19. Yoon, J. W. *et al.* Time-asymmetric loop around an exceptional point over the full optical communications band. *Nature* **562**, 86-90, doi:10.1038/s41586-018-0523-2 (2018).

20. Hokmabadi, M. P., Schumer, A., Christodoulides, D. N. & Khajavikhan, M. Non-Hermitian ring laser gyroscopes with enhanced Sagnac sensitivity. *Nature* **576**, 70-74, doi:10.1038/s41586-019-1780-4 (2019).

21. Lai, Y.-H., Lu, Y.-K., Suh, M.-G., Yuan, Z. & Vahala, K. Observation of the exceptional-point-enhanced Sagnac effect. *Nature* **576**, 65-69, doi:10.1038/s41586-019-1777-z (2019).

22. Miri, M.-A. & Alù, A. Exceptional points in optics and photonics. *Science* **363**, eaar7709, doi:10.1126/science.aar7709 (2019).

23. Muniz, A. L. M. *et al.* 2D Solitons in PT-Symmetric Photonic Lattices. *Physical Review Letters* **123**, 253903, doi:10.1103/PhysRevLett.123.253903 (2019).

24. Xia, S. *et al.* Nonlinear tuning of PT symmetry and non-Hermitian topological states. *Science* **372**, 72-76, doi:10.1126/science.abf6873 (2021).

25. Ding, K., Ma, G., Zhang, Z. Q. & Chan, C. T. Experimental Demonstration of an Anisotropic Exceptional Point. *Physical Review Letters* **121**, 085702, doi:10.1103/PhysRevLett.121.085702 (2018).

26. Tang, W. *et al.* Exceptional nexus with a hybrid topological invariant. *Science* **370**, 1077-1080, doi:10.1126/science.abd8872 (2020).

27. Li, Y. *et al.* Anti–parity-time symmetry in diffusive systems. *Science* **364**, 170-173, doi:10.1126/science.aaw6259 (2019).

28. Dong, Z., Li, Z., Yang, F., Qiu, C.-W. & Ho, J. S. Sensitive readout of implantable microsensors using a wireless system locked to an exceptional point. *Nature Electronics* **2**, 335-342, doi:10.1038/s41928-019-0284-4 (2019).

29. Shao, L. *et al.* Non-reciprocal transmission of microwave acoustic waves in nonlinear parity–time symmetric resonators. *Nature Electronics* **3**, 267-272, doi:10.1038/s41928-020-0414-z (2020).

30. Klauck, F. *et al.* Observation of PT-symmetric quantum interference. *Nature Photonics*, doi:10.1038/s41566-019-0517-0 (2019).

31. Dembowski, C. *et al.* Encircling an exceptional point. *Physical Review E* **69**, 056216, doi:10.1103/PhysRevE.69.056216 (2004).

32. Choi, Y., Hahn, C., Yoon, J. W., Song, S. H. & Berini, P. Extremely broadband,

on-chip optical nonreciprocity enabled by mimicking nonlinear anti-adiabatic quantum jumps near exceptional points. *Nat Commun* **8**, 14154, doi:10.1038/ncomms14154 (2017).
33. Zhang, X.-L., Song, J.-F., Chan, C. T. & Sun, H.-B. Distinct outcomes by dynamically encircling an exceptional point along homotopic loops. *Physical Review A* **99**, 063831, doi:10.1103/PhysRevA.99.063831 (2019).
34. Lopez-Galmiche, G. *et al.* in *Conference on Lasers and Electro-Optics.* FM1A.3 (Optical Society of America).
35. Gao, T. *et al.* Observation of non-Hermitian degeneracies in a chaotic exciton-polariton billiard. *Nature* **526**, 554-558, doi:10.1038/nature15522 (2015).
36. Berry, M. V. Quantal phase factors accompanying adiabatic changes. *Proceedings of the Royal Society of London. A. Mathematical and Physical Sciences* **392**, 45-57, doi:doi:10.1098/rspa.1984.0023 (1984).
37. Hassan, A. U. *et al.* Chiral state conversion without encircling an exceptional point. *Physical Review A* **96**, doi:10.1103/PhysRevA.96.052129 (2017).
38. Luo, L.-W. *et al.* WDM-compatible mode-division multiplexing on a silicon chip. *Nature Communications* **5**, 3069, doi:10.1038/ncomms4069 (2014).
39. Khurgin, J. B. *et al.* Emulating exceptional-point encirclements using imperfect (leaky) photonic components: asymmetric mode-switching and omni-polarizer action. *Optica* **8**, 563-569, doi:10.1364/OPTICA.412981 (2021).
40. Aleshkina, S. S. *et al.* All-fiber polarization-maintaining mode-locked laser operated at 980 nm. *Optics Letters* **45**, 2275-2278, doi:10.1364/OL.391193 (2020).